\documentclass[12pt]{article}

\usepackage{amsmath}
\usepackage{amssymb}
\usepackage{dsfont} 

\usepackage{graphicx}

\usepackage{color}

\newcommand{\be}{\begin{equation}}
\newcommand{\ee}{\end{equation}}
\newcommand{\bel}[1]{\begin{equation}\label{#1}}
\newcommand{\bea}{\begin{eqnarray}}
\newcommand{\eea}{\end{eqnarray}}
\newcommand{\balign}{\begin{align}}
\newcommand{\ealign}{\end{align}}
\newcommand{\ba}{\begin{array}}
\newcommand{\ea}{\end{array}}
\newcommand{\bfig}{\begin{figure}}
\newcommand{\efig}{\end{figure}}

\newcommand{\eref}[1]{(\ref{#1})}

\allowdisplaybreaks

\newcommand{\exval}[1]{\mbox{$\langle \, {#1}\, \rangle$}}

\newcommand{\pdt}{\frac{\partial}{\partial t}}
\newcommand{\pdx}{\frac{\partial}{\partial x}}

\newcommand{\rmd}{\mathrm{d}}
\newcommand{\rme}{\mathrm{e}}

\newcommand{\diag}{\mathrm{diag}}

\newcommand{\N}{{\mathbb N}}

\def\P{\Phi}
\begin{document}

\title{KPZ modes in $d$-dimensional directed polymers}
\author{G.M.~Sch\"utz$^{1}$ 
\and B. Wehefritz--Kaufmann $^{2}$ 
}

\maketitle

{\small
\smallskip
\noindent $^{~1}$Institute of Complex Systems II,
Forschungszentrum J\"ulich, 52425 J\"ulich, Germany
\\
\noindent Email: g.schuetz@fz-juelich.de

\smallskip 
\noindent $^{2}$ Purdue University, Department of Mathematics and Department of Physics and Astronomy, 150 N. University Street. West Lafayette, IN 47906, USA\\
\noindent Email: ebkaufma@math.purdue.edu
}

\begin{abstract}
We define a stochastic lattice model for a fluctuating directed polymer in 
$d\geq 2$ dimensions. This model can be alternatively interpreted as a 
fluctuating random path in 2 dimensions, or a one-dimensional asymmetric 
simple exclusion process with $d-1$ conserved species of particles. The 
deterministic large dynamics of the directed polymer are shown to be given 
by a system of coupled Kardar-Parisi-Zhang (KPZ) equations and diffusion 
equations. Using non-linear fluctuating hydrodynamics and mode coupling theory 
we argue that stationary fluctuations in any dimension $d$ can only be of 
KPZ type or diffusive. The modes are pure in the sense that there are only 
subleading couplings to other modes, thus excluding the occurrence of modified 
KPZ-fluctuations or L\'evy-type fluctuations which are common for more than one
conservation law. The mode-coupling matrices are shown to satisfy the 
so-called trilinear condition.
 \\[.3cm]\textbf{Keywords:} Directed polymer, Exclusion process, 
KPZ equation, Non-linear fluctuating hydrodynamics

\end{abstract}

\newpage

\section{Introduction}
\label{s_intro}

The dynamics of one-dimensional many-body systems is presently a topic of 
intense study. One of the main motivations is to study anomalous transport 
phenomena which arise in different contexts and various physical scenarios
even when interactions are short-ranged. Specific topics of interest are one-dimensional stochastic equations with local conservations laws (in particular 
for interface dynamics in the universality class of the one-dimensional noisy 
Kardar-Parisi-Zhang (KPZ) equation) and stationary spatio-temporal fluctuations 
in driven diffusion, or anharmonic chains or Hamiltonian fluid dynamics, 
see e.g. the collection of articles in \cite{Lepr16} and in the first issue
of volume 160 of the Journal of Statistical Physics (2015) for recent overviews.
In the case of a single locally conserved quantity the long wave-length 
fluctuations of the conserved field are generally either diffusive with 
dynamical exponent $z=2$ or in the KPZ universality class \cite{Halp15}
with dynamical exponent $z=3/2$.

In this article, we will focus on \emph{coupled} one-dimensional stochastic 
equations with more than one conservation law. They show a much richer behavior 
than the single KPZ equation, depending on the details of the models. 
Fluctuations of the conserved fields can be in a modified KPZ universality 
class \cite{Spoh15} or, more intriguingly, in a discrete family of L\'evy 
universality classes \cite{Popk15b} where the dynamical exponents $z_i$ are 
the Kepler ratios of neighbouring Fibonacci numbers and the universal 
scaling forms of the dynamical structure function are $z_i$-stable 
L\'evy distributions. The first member in this family 
is a mode with dynamical exponent $z=3/2$ as in KPZ, but L\'evy scaling function 
which very recently was proved rigorously for energy fluctuations in a harmonic
chain with energy-conserving noise \cite{Bern16}. The second member 
with dynamical exponent $z=5/3$ was first firmly established using 
mode coupling theory for the heat mode in Hamiltonian dynamics for a 
one-dimensional fluid \cite{vanB12}. Also the limiting value of the Kepler 
ratios, which is the famous golden mean, can arise \cite{Spoh15,Popk15b,Popk15a}.

Here we address the nature of the dynamical structure functions in a higher 
dimensional setting, viz. for the contour fluctuations in a lattice 
model for a directed polymer in $d\geq 2$ dimensions, somewhat
in the spirit of the space-continuous polymer model of \cite{Erda93} 
for $d=3$. Our lattice model can be mapped to a fluctuating random 
path in two dimensions and also to a one-dimensional exclusion process 
\cite{Spoh91,Ligg99,Schu01,Scha10} generalized to $d-1$ species of particles. 
We use the latter mapping, taking two different approaches to study the 
large-scale dynamics and the spatio-temporal fluctuations in the stationary state.

First, focussing on $d=3$, a dynamical mean-field approach for the
particle densities leads to a system of 
two coupled partial differential equations that each look like a Burgers 
equation. By introducing a generalized height variable, these equations
become coupled KPZ equations. The couplings depend 
on the rates of the original exclusion process. By varying the rates, one can 
systematically study the different universality classes. However, two of the 
entries in the coupling matrices will always remain zero, regardless of the rates 
in the underlying exchange process.

The second approach is based on nonlinear fluctuating hydrodynamics,
which has emerged as a widely applicable and powerful tool for the study 
of stationary fluctuations of the locally conserved quantities such as energy, 
momentum, or particle densities \cite{Spoh14}. From the exact current-density
relation we compute the mode-coupling matrices which allow us to deduce the
dynamical universality classes that can occur in the model in any dimension
$d\geq 2$. We find that only KPZ modes and diffusive modes may occur and that
these modes have only subleading couplings between them, which excludes also
the occurrence of the modified KPZ universality class. We point out that the 
mode coupling matrices satisfy the so-called trilinear condition 
which is relevant for the Gaussian nature of the
invariant measure of the associated coarse-grained system of 
coupled noisy KPZ-equations \cite{Ferr13,Funa15}.

This paper is organized in the following way. We start by defining  in Section 2
the directed polymer model in any dimension $d$ that is a generalization of the 
well-known correspondence between the single--species asymmetric diffusion model 
and a growing and fluctuating interface in $d=2$. In Section 3 we focus on
$d=3$ and first derive a system of 2 coupled non-linear partial differential equations for a generalized height function from a coarse-graining of the model. 
Next we study fluctuations via nonlinear fluctuating hydrodynamics. Chapter 4
contains a calculation of the mode--coupling coefficients for an 
$n$-component particle exchange process, corresponding to a directed polymer
in $d=n+1$ dimensions. Discussing the case $n=2$ in detail yields a direct comparison with the height model results. In Section 5 we summarize our results and point to
some open problems.

\section{Directed polymer in $d$ dimensions, generalized height function, and the multi-species ASEP}

There is a very nice and well-known mapping between the one-dimensional
single-species asymmetric simple exclusion process (ASEP) and a growing and 
fluctuating interface on a two-dimensional substrate \cite{Meakin, Plischke}. 
The contour of this interface can equally be interpreted as a model for a
directed polymer living on a square lattice in two dimensions. The conformation 
of the polymer, or equivalently, the height function of the interface, is given 
by a microstate of the ASEP. 

Generalizing to multi-species simple exclusion processes \cite{Schu03}, it is 
natural to search for an analogous construction in higher dimensions. We 
demonstrate that there is indeed a natural way of defining a directed polymer 
model in any dimension. This is achieved by identifying the directed polymer 
with a directed path on a plane perpendicular to the $(1,1,\dots,1)$-direction
of a hypercubic lattice
and introducing 
an associated generalized height function. Below we present the details of this 
mapping and show that by deriving an equation for the time evolution of 
the height variable one obtains a set of coupled differential equations that 
describe either diffusive or KPZ or mixed behavior. The same equations can be 
derived from the corresponding multispecies simple exclusion process and its 
master equation dynamics.

\subsection{Details}

Consider $d$ species of particles with exclusion, i.e., at most one particle 
per site, on a one-dimensional chain of $L$ sites, counting a "vacancy" as 
a species. Particles of different species $\alpha$ and $\beta$ randomly 
interchange their positions with rates $g_{\alpha,\beta}$ see Sec. \eref{PEP} 
for a precise definition of this multi-species exclusion process.
Then each configuration of the chain can be mapped to a directed path on a 
$d$-dimensional hypercubic lattice which is later projected onto a plane
perpendicular to  the $(1,1,\dots,1)$-direction: 
As you step along the chain, the corresponding steps of the path on the hypercube 
are given by what species of particle you pass, with each species corresponding 
to one of the $d$ basis vectors of the hypercube with unit length $a$. 
Thus each step increases the height of the corresponding segment of the 
directed polymer by $a/\sqrt{d}$ above its anchor point. By convention we 
take the anchor point to be the origin $\vec{0}=(0,0,\dots,0)$.
We assume no external potential so that in the stationary state each 
conformation of the directed polymer is equally likely.

For a hypercube with unit lattice constant $a=1$ the contour length of the
polymer is $Ld$. The endpoint of the polymer after the $L$ steps of the
underlying particle configuration is at height $L/\sqrt{d}$. Its position
is determined by the (conserved) number of particles of each 
species in the chain. In particular, if the number $N_\alpha$ of particles is the same
for each species $\alpha$, i.e., if $N_\alpha=L/d$ 
then the endpoint of the polymer has coordinates 
$L/\sqrt{d}(1,1,\dots,1)$. The projection of the position of the polymer 
after $k$ steps along the chain onto the hyperplane perpendicular to the
$(1,1,\dots,1)$-direction defines a generalized height variable which
is a $d-1$-dimensional vector.

\subsection{Example in $d=3$, leading to a path in $d=2$}

For definiteness we discuss in more detail the case $d=3$, where our generalized 
height will be shown as the position of the path projected onto a plane 
perpendicular to the (111) axis of the cube. This path will be in two dimensions. 
The dynamics of the system is then represented by elementary moves of this path, 
where one site along the path moves in the only way which is determined by
the constraints imposed by the particle exchange dynamics of the exclusion
process with three conserved particle species and no vacant sites. Notice
that since $L$ is fixed by the dynamics, the particle exchange dynamics 
correspond to only two genuine conservation laws. This can be seen by identifying
one species with vacant sites. 
We consider periodic boundary conditions for the exclusion
process with an equal number of particles of each species
which corresponds to periodic boundary conditions for the directed polymer.

To be concrete, we start from a two--species asymmetric exclusion model 
on a ring with $L$ sites where each site $k$ is either empty or occupied 
by at most one particle of type $A$ or $B$. For our purposes
it is convenient to think of a vacancy as a further species of particles,
denoted by $\P$. 
A microscopic particle configuration is specified by an
array of $L$ symbols $X_k$ where $X_k\in \{\P,A,B\}$, or, equivalently,
by occupation numbers $n_k^X = \delta_{X,X_k}$ which are equal to
1 if the particle at site $k$ is of type $X$ and zero otherwise.
It defines a conformation of the directed polymer as described above.

The Markovian stochastic dynamics consists of nearest-neighbour 
particle exchanges $X_i,X_{i+1} \to X_{i+1}, X_i$ as follows:
\be
\ba{ll}
\mbox{Transition} & \;\; \mbox{Rate} \\[2mm]
A\,B \rightarrow B\,A & \;\; r_1 \\
B\,A \rightarrow A\,B & \;\;r_2 \\
A\,\P \rightarrow \P\,A & \;\;r_3 \\
\P\,A \rightarrow A\,\P & \;\;r_4 \\
B\,\P \rightarrow \P\,B & \;\;r_5 \\
\P\,B \rightarrow B\,\P & \;\;r_6
\ea
\ee
In order to ensure equal equilibrium probabilities for all conformations of the
polymer (corresponding to the uniform measure for particle configurations)
we impose pairwise balance \cite{Schu96} which yields
\begin{equation}
\label{pairbal}
r_1+r_4+r_5=r_2+r_3+r_6.
\end{equation}
The uniform distribution leads to a complete absence of stationary correlations
in the thermodynamic limit $L\to\infty$.

The link with the height function and the two-dimensional
random path is established as follows.
With each of the three species ($A$, $B$ or vacancy $\Phi$)
we associate one of the three canonical basis vectors $\vec{e}_i$
of the 3-$d$ cubic lattice. Thus, starting from the anchor point 
of the polymer (say, the origin $\vec{0} = (0,0,0)$),
the height along the $(1,1,1)$ axis is $k/\sqrt{3}$ where $k$ is the
lattice site of the one-dimensional chain of particles. 
The particle configuration from site 1 to site $k$ on the chain 
then describes the position of the height vector $\vec{H}_k=\sum_{j=1}^k 
\vec{e}^{X_j}_j$ in the plane perpendicular to the $(111)$ direction, reflecting
the position of the polymer in three dimensional space at height 
$k/\sqrt{3}$.

\begin{figure}[ht]
\begin{center}
\includegraphics[width=.3\textwidth]{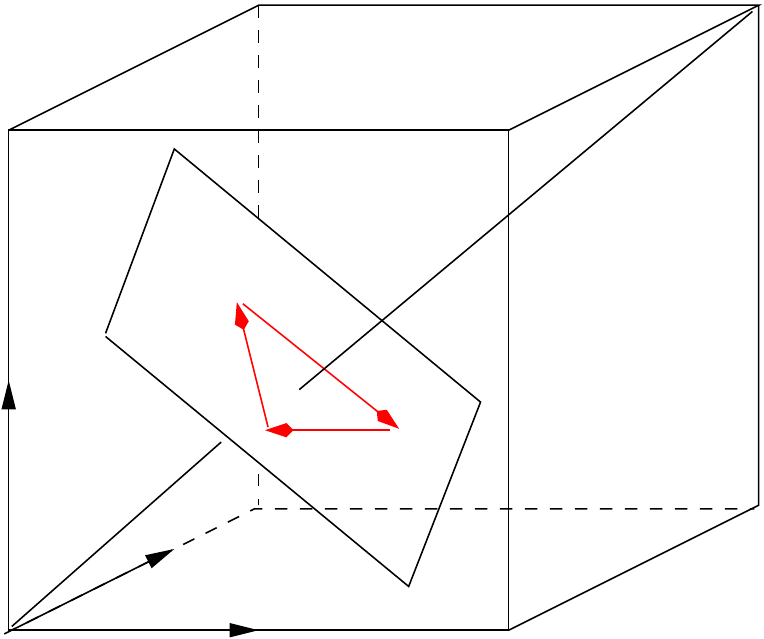} \hspace*{2cm}
\includegraphics[width=.3\textwidth]{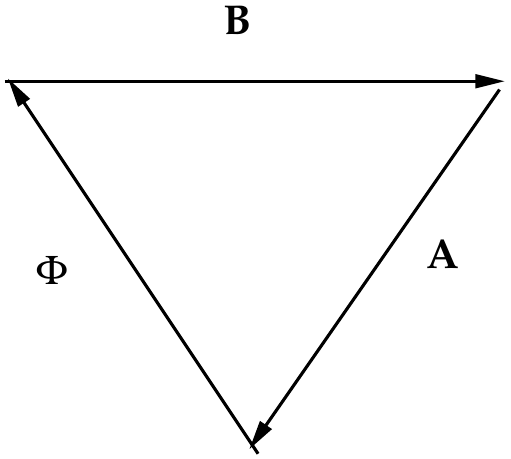}
\end{center}
\caption {Projection of the position of the directed polymer 
onto the plane perpendicular to the $(1,1,1)$ direction (left) and
directions for the 2D height function (right).}
\label{cube}
\end{figure}

The projection of the three basis vectors onto the plane perpendicular to 
the $(1,1,1)$ direction is shown in Fig. \ref{cube} (left).
This results in the following three normalized vectors:
$$
\vec{v_{\P}}= \frac{1}{\sqrt{6}} \left( \begin{matrix} 2\\ -1 \\-1\end{matrix}\right),\;\;\;\;\;\;
\vec{v_A}= \frac{1}{\sqrt{6}} \left( \begin{matrix} -1\\ 2 \\-1\end{matrix}\right),\;\;\;\;\;\;
\vec{v_B}= \frac{1}{\sqrt{6}} \left( \begin{matrix}-1\\ -1\\2\end{matrix}\right).
$$
Now we can pick basis vectors for the plane perpendicular to the $(1,1,1)$ direction, e.\  g.\
$$
\vec{b_1}= \frac{1}{\sqrt{6}} \left( \begin{matrix} -1\\ -1 \\2\end{matrix}\right),\;\;\;\;\;\;
\vec{b_2}= \frac{1}{\sqrt{2}} \left( \begin{matrix} 1\\ -1 \\0\end{matrix}\right)
$$
Expressing the vectors $\vec{v_i}; i=\P,A,B$ in terms of the two basis vectors $\vec{b_1}$ and $\vec{b_2}$, 
they become the
two-dimensional unit vectors in the projection plane (see Fig.
 \ref{cube} (right)):
\begin{equation}
\vec{\P}=\left( \begin{matrix}-\frac{1}{2}\\ \frac{\sqrt{3}}{2}\end{matrix}\right),\;\;\;\;\;\;
\vec{A}=\left( \begin{matrix}-\frac{1}{2}\\ -\frac{\sqrt{3}}{2}\end{matrix}\right),\;\;\;\;\;\;
\vec{B}=\left( \begin{matrix}1\\ 0\end{matrix}\right)
\end{equation}

The projected height vector  at level $k$, which is 
the generalized height function we are after, is then given by
\be 
\vec{H}^\bot_k = \vec{H}^\bot_0 + 
\sum_{j=1}^k \left( n_k^\P \vec{\P} + n_k^A \vec{A} + n_k^B \vec{B} \right)
\ee
where $\vec{H}^\bot_0$ is the reference point (taken to be the origin
in the description above). This shows that the local occupation numbers
give the (discrete) height gradient 
\bel{discretegradient}
\nabla_{(111)} \vec{H}^\bot_k := \vec{H}^\bot_k - \vec{H}^\bot_{k-1}
= n_k^\P \vec{\P} + n_k^A \vec{A} + n_k^B \vec{B}
\ee
in $(111)$-direction. Fluctuations in the height vector are described by
nearest neighbour particle swaps as defined above. 

\begin{figure}[ht]
\includegraphics[width=.4\textwidth]{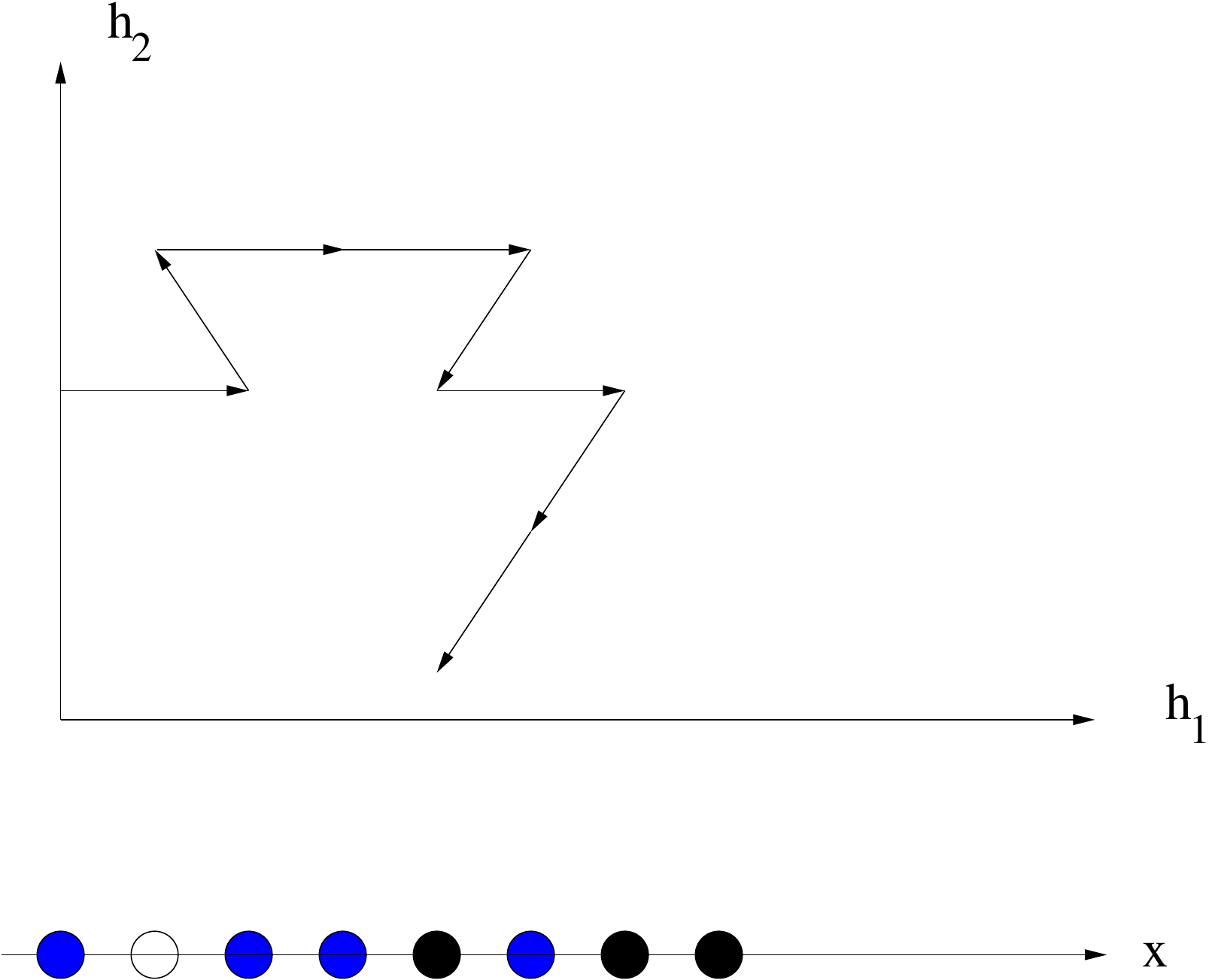}
\hspace{1cm}
\includegraphics[width=.4\textwidth]{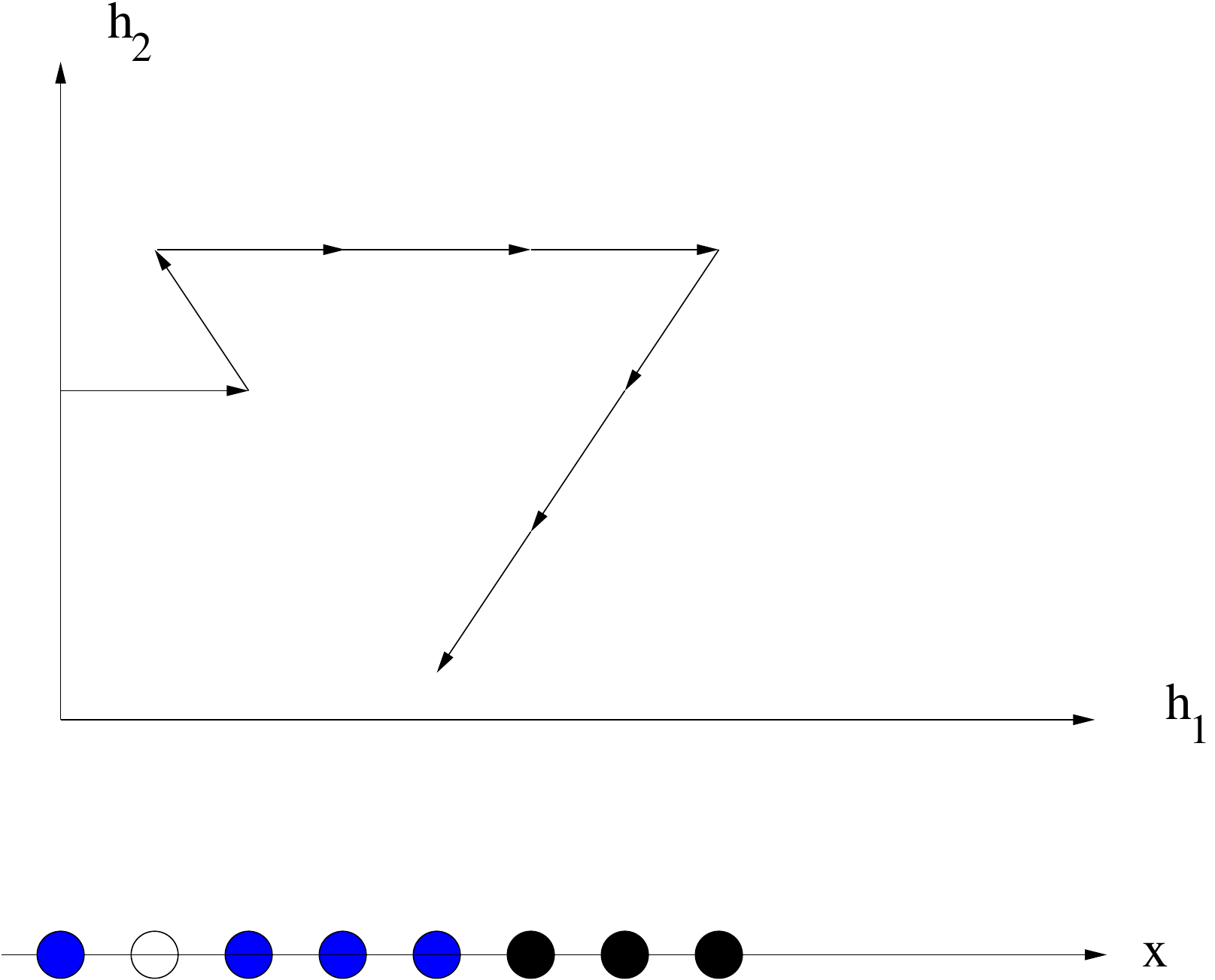}
\caption {Two-dimensional random path on the honeycomb lattice and diffusing 
particles. From the left picture to the right picture, the black particle on 
lattice site 5 and the blue particle on lattice site 6 have interchanged 
places and the path has changed accordingly.}
\label{figHeightmodel}
\end{figure}

Correspondingly the surface path in the plane perpendicular to the $(111)$ direction becomes
a planar random path on a honeycomb lattice with unit lattice constant 
(Fig. \ref{figHeightmodel}). A change in the path happens when
two particles interchange places.

\section{Coarse-grained dynamics and stationary spatio-temporal fluctuations}

\subsection{Coupled KPZ equations for the height function}

In order to study the large-scale behaviour of the height function for
arbitrary initial states we define a coarse-grained 
two-dimensional height variable 
\be
\vec{h}(x,t) =\left( \begin{matrix}h_1 (x,t)\\ h_2 (x,t)\end{matrix}\right).
\ee
Since $\vec{A}+\vec{B}+\vec{\P}=0$, it follows that $\rho_A=\rho_B=\rho_{\P}=1/3$ for the average particle densities. 
Therefore
we define coarse-grained local densities $\rho_A(x,t), \rho_B(x,t)$ and $\rho_{\P}(x,t)$ 
for species $A,B$ and $\Phi$, respectively, as follows:
\begin{eqnarray}
\rho_A(x,t)&=& \frac{1}{3}+\frac{2}{3} \;(\vec{A}\cdot\nabla\vec{h}(x,t))\nonumber\\
\rho_B(x,t)&=& \frac{1}{3}+\frac{2}{3} \;(\vec{B}\cdot\nabla\vec{h}(x,t))\nonumber\\
\rho_{\P}(x,t)&=& \frac{1}{3}+\frac{2}{3} \;(\vec{\P}\cdot\nabla\vec{h}(x,t))
\end{eqnarray}
Here, $\nabla \vec{h}$ denotes the one-dimensional derivative in the 
direction of the diffusing particles, i.e., along the coarse-grained
chain in $(111)$-direction.
Each of the densities fluctuates around its equilibrium value $\frac{1}{3}$ and will be changed proportionally to the  change $\nabla\vec{h}$ in the height variable $\vec{h}(x,t)$ projected onto the respective growth direction.

In order to derive a nonlinear evolution equation for $\vec{h}(x,t)$
we recall that the local particle density describes the gradient of
the height vector, see \eref{discretegradient} for the discrete case.
In order to obtain an equivalent continuum description
we symmetrize the discrete gradient and
expand the $\rho_i$ for $i=A,B,\P$ around $x$ to second order, 
leading to
\begin{equation}
\rho_A(x\pm\frac{1}{2})\simeq \frac{1}{3}+\frac{2}{3} \;(\vec{A}\cdot\nabla\vec{h})\mp\frac{1}{3}( \vec{A} \cdot \Delta \vec{h})+\cdots
\end{equation}
The next step is to consider the time evolution of the height variable 
$\vec{h}(x,t)$.
From the absence of correlations in the stationary distribution and the
dynamical rules of the model we find 
\begin{eqnarray}
\label{diffeqgeneral}
\dot{\vec{h}}(x)&=&(r_3-r_4)(\vec{\Phi}-\vec{A})\;\rho_{\Phi}(x-\frac{1}{2})\;\rho_A(x+\frac{1}{2})\nonumber\\
&&+(r_1-r_2)(\vec{B}-\vec{A})\;\rho_B (x-\frac{1}{2})\;\rho_A(x+\frac{1}{2})\nonumber\\
&&+(r_5-r_6)(\vec{\P}-\vec{B})\;\rho_{\P} (x-\frac{1}{2})\;\rho_B(x+\frac{1}{2}).
\end{eqnarray}
This equation describes how the height variable will change after two particles on the lattice will have interchanged places. The increase is proportional to the density of particles and is proportional to the growth direction associated with the interchange process. 

We will adopt the following notation:
\begin{eqnarray}
 r_1+r_2&=&p \\
 r_5+r_6&=&q \\
r_3-r_4 &=& f_1\\
r_5 - r_6 &=& f_2\\
r_1-r_2 &=& f_1-f_2
\end{eqnarray}
The last equation follows from the pairwise balance requirement 
Eq.~ (\ref{pairbal}).
Putting everything into Eq. \eref{diffeqgeneral} and denoting
transposition of a vector or matrix by a superscript $T$, we obtain 
\begin{eqnarray}
\label{gen_eq}
\pdt \vec{h}(x)&=&\left( \begin{matrix}\frac{1}{6} (f_1-2f_2)\\ \frac{1}{2\sqrt{3}}f_1\end{matrix}\right)+
\left( \begin{matrix}\frac{(f_1-2f_2)}{6} &- \frac{f_1}{2\sqrt{3}}\\- \frac{f_1}{2\sqrt{3}} & -\frac{(f_1-2f_2)}{6} \end{matrix}\right) \nabla \vec{h}\nonumber\\
&&
+\left( \begin{matrix}\frac{(p+q)}{4} &\frac{(p-q)}{4\sqrt{3}}\\ \frac{(p-q)}{4\sqrt{3}} & \frac{2r_1+2r_5+3(p+q)}{12} \end{matrix}\right) \nabla^2 \vec{h}
 \nonumber\\&&
+\left( \begin{matrix}(\nabla \vec{h})^T\left( \begin{matrix}-\frac{(f_1-2f_2)}{3} &- \frac{f_1}{2\sqrt{3}}\\- \frac{f_1}{2\sqrt{3}} &0 \end{matrix}\right)\nabla \vec{h}\\
 (\nabla \vec{h})^T\left( \begin{matrix}0&- \frac{(f_1-2f_2)}{6} \\- \frac{(f_1-2f_2)}{6}  & -\frac{f_1}{\sqrt{3}} \end{matrix}\right)\nabla \vec{h}\end{matrix}\right)+\cdots
\end{eqnarray}
Taking the gradient on both sides one recognizes two coupled KPZ equations
with non-vanishing drift term which mixes the two height components.

We express this system of non-linear coupled equations in terms of eigenfunctions
of matrix multiplying $\nabla \vec{h}$. The eigenvalues are 
\begin{equation}
\label{lambda12}
v_{1,2}=\pm \frac{1}{3}\sqrt{f_1^2-f_1 f_2+f_2^2} =: \pm \frac{1}{3} s(f_1,f_2).
\end{equation}
The expression under the square root is always positive except for $f_1=f_2=0$ 
in which case not only $v_1=v_2=0$ but where also the non-linear term vanishes. 
This corresponds to the (boring) case of symmetric diffusion which we exclude 
from our considerations.
It is interesting that the two eigenvalues $\lambda_{1,2}$ are then never
equal. This implies that the drift term cannot be removed
by a Galilei transformation.

When $f_1=0$ we do not need to apply a similarity transformation. 
The result of the transformation to eigenmodes 
$\vec{\tilde{h}}(x)$ for $f_1\neq 0$ is
\begin{eqnarray}
\label{gen_eq_diag}
\pdt \vec{\tilde{h}}(x)&=&\left( \begin{matrix}\tilde{v}_1\\
\tilde{v}_2\end{matrix}\right)+
\left( \begin{matrix}- \frac{1}{3}s(f_{1},f_{2}) &0\\
0 & \frac{1}{3} s(f_1,f_2) \end{matrix}\right) \nabla \vec{\tilde{h}}(x)\nonumber\\
&&
+\left(  \begin{matrix} M_{11} & M_{12}\\M_{21} & M_{22}\end{matrix} \right) \nabla^2 \vec{\tilde{h}}(x)
+\left( \begin{matrix}(\nabla \vec{\tilde{h}}(x))^T\left( \begin{matrix}N_{11} &N_{12}\\N_{21}&0 \end{matrix}\right)\nabla \vec{\tilde{h}}(x)\\
 (\nabla \vec{\tilde{h}}(x))^T\left( \begin{matrix}0&P_{12} \\ P_{21} & P_{22} \end{matrix}\right)\nabla \vec{\tilde{h}}(x)\end{matrix}\right)+\cdots
\end{eqnarray}
with  the average growth velocities
{\footnotesize
\begin{eqnarray}
\tilde{v}_1 &=&\frac{f_1(f_1-2f_2+s(f_1,f_2))(2f_1^2+2f_2(f_2+s(f_1,f_2))-f_1(2f_2+s(f_1,f_2))}{6 s(f_1,f_2)|f_1| }\nonumber \\ 
\tilde{v}_2 &=& \frac{f_1(-f_1+2f_2+s(f_1,f_2))(2f_1^2+2f_2(f_2-s(f_1,f_2))+f_1(-2f_2+s(f_1,f_2))}{6 s(f_1,f_2)|f_1| }\nonumber 
\end{eqnarray}
} of the two projected height variables in normal mode coordinates
and the matrix elements
{\footnotesize
\begin{eqnarray}
M_{11} &=&  \frac{f_2(3(p - q) -( r_1 + r_5)) + (f_1-f_2)(3(p - q) + r_1 + r_5) + 2 s(f_1,f_2)(3(p + q) + r_1 + r_5)}{24 s(f_1,f_2)}\nonumber \\
   M_{22}&=&  \frac{ -f_2(3(p -q) + r_1 + r_5) - (f_1-f_2)(3(p - q) + r_1 + r_5) + 2 s(f_1,f_2) (3(p + q) + r_1 + r_5)}{24 s(f_1,f_2)} \nonumber \\
M_{12}&=& -\frac{(-f_1+2 f_2 + 2s(f_1,f_2))\sqrt{6f_1^2 -6 f_1 f_2 +f_2(2 f_2-s(f_1,f_2)) }
    }{24f_1s(f_1,f_2)
    \sqrt{6f_1^2 -6 f_1 f_2 +f_2(2 f_2+s(f_1,f_2)) } }\times   \nonumber\\
    && \times  (-2f_1 (r_1+r_5)+f_2(-p + q + (r_1 + r_5)) )\nonumber\\
    M_{21}&=& -\frac{(f_1-2f_2 + 2s(f_1,f_2))\sqrt{6f_1^2 -6 f_1 f_2 +f_2(2 f_2+s(f_1,f_2))}
    }{24f_1s(f_1,f_2)
    \sqrt{6f_1^2 -6 f_1 f_2 +f_2(2 f_2-s(f_1,f_2)) } }\times   \nonumber\\
    && \times  (-2f_1 (r_1+r_5)+f_2(-p + q + (r_1 + r_5)))\nonumber\\
    N_{11}&=& -\frac{f_1(f_1-2f_2+s(f_1,f_2)) \sqrt{6f_1^2 -6 f_1 f_2 +f_2(2 f_2+s(f_1,f_2))}    }{
  3s(f_1,f_2)|f_1|}\nonumber\\
    P_{22}&=& -\frac{f_1(-f_1+2f_2+s(f_1,f_2)) \sqrt{6f_1^2 -6 f_1 f_2 +f_2(2 f_2-s(f_1,f_2)) }    }{
  3s(f_1,f_2)|f_1|}\nonumber\\
  N_{12}&=&N_{21}=-\frac{|f_1|((f_1-f_2)(-(f_1-f_2)^2+5(f_1-f_2)s(f_1,f_2))+f_1^2(-f_1+s(f_1,f_2)))   }{6 s(f_1,f_2)(f_1)\sqrt{6f_1^2 -6 f_1 f_2 +f_2(2 f_2-s(f_1,f_2)) }}\nonumber\\ 
     P_{12}&=&P_{21}=-\frac{|f_1|((f_1-f_2)((f_1-f_2)^2+5(f_1-f_2)s(f_1,f_2))+f_1^2(f_1+s(f_1,f_2)))   }{6 s(f_1,f_2)(f_1)\sqrt{6f_1^2 -6 f_1 f_2 +f_2(2 f_2+s(f_1,f_2)) }}\nonumber
\end{eqnarray}
}
of the phenomenological diffusion matrix.
The matrices $N$ and $P$ are the mode coupling matrices which yield the
structure of the non-linear part of the coarse-grained evolution equation.
We checked that for $f_1\neq 0$ the expressions under the square roots will always be 
positive or zero, and that the denominators are not zero. By rewriting these
equations in terms of the height gradients $\vec{\tilde{\rho}}(x,t) = 
\nabla \vec{\tilde{h}}(x,t)$ one gets a system of coupled Burgers equations.

\subsection{Stationary space-time fluctuations}

As has become clear from the previous section it is convenient
to work with height gradients which map to densities $\rho_\alpha(x,t)$
which are globally conserved, i.e., 
$\int \rmd x \rho_\alpha(x,t) = L \rho_\alpha$
for a system of length $L$.
A fundamental quantity of interest is the dynamical structure function
which are the stationary two-time correlations of the height gradients.
For $n$ conserved densities this is an $n\times n$ matrix with the
two-point correlations between the (centered) densities 
$u_\alpha(x,t)
= \rho_\alpha(x,t) -\rho_\alpha$ at time $t_0$ and $t_0+t$. Because of
stationarity $t_0$ is immaterial and can be set to 0.

\subsubsection{Nonlinear fluctuating hydrodynamics}
\label{NFH}

In order to study such a system with noisy dynamics on a coarse-grained
level we follow the powerful and nonlinear fluctuating hydrodynamics
(NLFH) approach \cite{Spoh14} whose essence and main insights 
we briefly summarize. 

Consider a system with $n$ conserved densities $\rho_\alpha$ 
and associated locally conserved currents
$j_\alpha$. On coarse-grained 
Eulerian scale, where the noise drops out as a result
of the law of large numbers, the conservation laws imply that the
densities satisfy the nonlinear system of PDE's \cite{Spoh91,Kipn99}
\bel{hydro}
\pdt \vec{\rho}(x,t) + \pdx \vec{j}(x,t) = 0
\ee
where component $\rho_\alpha(x,t)$ of the vector $\vec{\rho}(x,t)$ is a
coarse-grained conserved quantity and the component $j_\alpha(x,t)$ 
of the current vector $\vec{j}(x,t)$ is the associated locally conserved current. 
Notice that in our convention $\vec{\rho}$ and  $\vec{j}$ are regarded as 
column vectors. 

Because of local stationarity under Eulerian scaling
the current is a function of $x$ and $t$ only through its dependence 
on the local conserved densities.
Hence these equations can be rewritten as
\begin{equation}
\label{hyper}
\frac{\partial}{\partial t} \vec{\rho}(x,t) + 
J(x,t) \frac{\partial}{\partial x} \vec{\rho}(x,t) = 0
\end{equation}
where $J(x,t)$ is the
current Jacobian with matrix elements $ J_{\alpha\beta} = 
\partial j_\alpha / \partial \rho_\beta$, understood as functions
of $x$ and $t$ via $\rho_\alpha(x,t)$ via the stationary current-density
relation $\vec{j}^\ast(\vec{\rho})$. In other words, 
$\vec{j}(x,t) = \vec{j}^\ast(\vec{\rho}(x,t))$.
Obviously, constant 
densities $\rho_\alpha$ are a (trivial) stationary solution of \eref{hyper}. 
Stationary fluctuations of the conserved quantities are captured in the
compressibility matrix $K$ that we shall not describe explicitly.

Up to this point the system \eref{hyper}, and therefore also its expansion in
$u_\alpha(x,t)$, is completely deterministic. In the NLFH
approach the effect of fluctuations is captured by adding a 
phenomenological diffusion matrix $D$ and white noise terms $\xi_i$.
This turns \eref{hyper} into a system of non-linear stochastic PDE's.
From renormalization group considerations it is known that polynomial 
non-linearities of order higher than 4 are irrelevant for the large-scale
behaviour and order 3 leads at most to logarithmic corrections if the
generic quadratic non-linearity is absent \cite{Devi92}. This justifies
an expansion to second order so that
the fluctuation fields $u_\alpha(x,t)$ satisfy the system of coupled
noisy Burgers equations
\begin{equation}
\label{coupledBurgers}
\partial_t \vec{u} =  - \partial_x \left( J \vec{u} + \frac{1}{2} \vec{u}^T \vec{H} \vec{u} - D \partial_x  \vec{u}
+ B \vec{\xi} \right)
\end{equation}
where $\vec{H}$ is a column vector whose entries $(\vec{H})_\alpha=H^{\alpha}$ are the Hessians
with matrix
elements $H^{\alpha}_{\beta\nu} = \partial^2 j_\alpha /(\partial \rho_\beta \partial \rho_\nu)$.
If the quadratic non-linearity is absent one has diffusive behaviour.
We stress that the Hessians $\vec{H}$ depend on the stationary densities
around which one expands, but not on space and time. Hence they
are fixed by the stationary current-density relation 
$\vec{j}^\ast(\vec{\rho})$.

In order to proceed further it is convenient to transform into
normal modes 
$\vec{\phi} =R \vec{u}$ where $ RJR^{-1} = \mathrm{diag}(v_\alpha)$ and the transformation matrix $R$.
The eigenvalues
$v_\alpha$ of $J$ play the role of characteristic speeds
 that
on microscopic scale describe the speed of local perturbations \cite{Popk03}.
One thus arrives at
\begin{equation}
\label{normalmodes}
\partial_t \phi_\alpha = -  \partial_x \left( v_\alpha \phi_\alpha +
\vec{\phi}^T G^{\alpha} \vec{\phi} - \partial_x  (\tilde{D} \vec{\phi})_\alpha
+ (\tilde{B} \vec{\xi})_\alpha \right)
\end{equation}
with $\tilde{D}=RDR^{-1}$ and $\tilde{B}=RB$. The matrices
\begin{equation}
\label{mcmatrix}
G^{\alpha} =  \frac{1}{2} \sum_\gamma R_{\alpha\gamma} (R^{-1})^T H^{\gamma} R^{-1}
\end{equation}
are the mode coupling matrices with the mode-coupling coefficients
$G^{\alpha}_{\beta\gamma}=G^{\alpha}_{\gamma\beta}$ which are, 
by construction, symmetric. They are said to satisfy the trilinear condition
if they satisfy also the symmetry 
$G^{\alpha}_{\beta\gamma}=G^{\beta}_{\alpha\gamma}$ \cite{Ferr13,Funa15}.

The main quantities of interest are then dynamical structure functions 
\bel{S-matrix}
S^{\alpha\beta}(x,t) = \exval{\phi^{\alpha}(x,t)\phi^{\beta}(0,0)}
\ee
which describe the stationary space-time fluctuations of the normal
modes. They satisfy the normalization
\bel{sfnorm}
\int_{-\infty}^\infty \rmd x \, S^{\alpha\beta}(x,t) = \delta_{\alpha,\beta} 
\ee
which arises from the conservation law and the normalization condition
$RKR^T = \mathds{1}$.
It is important to note that in the absence of long-range order and
long-range jumps generally
the product $JK$ of the Jacobian with the compressibility matrix $K$ is
symmetric, which can be proved mathematically rigorously \cite{Gris11}. 
This guarantees that on macroscopic scale the full non-linear
system (\ref{hyper}) is hyperbolic \cite{Toth03}, i.e., characteristic
velocities $v_\alpha$ are real. 

When the characteristic velocities are all 
different, i.e., in the strictly hyperbolic case, 
the off-diagonal terms $S^{\alpha\beta}$ decay quickly
and for long times
and large distances one is left with the diagonal elements
$S^{\alpha\alpha}(x,t)$ which are asymptotically universal functions
$S^{\alpha\alpha}(x,t) \sim t^{-1/z_\alpha} f (u_\alpha)$ with the scaling variable
$u_\alpha = (x-v_\alpha t)^{z_\alpha} /t$. Here $z_\alpha$ is the dynamical
exponent.

These scaling functions can be evaluated
using mode coupling theory \cite{Spoh14,Popk16}. 
As pointed out in the introduction, in systems
with short-range interactions
there is an infinite discrete family of universality classes
with dynamical exponents $z_\alpha$ that are the Kepler ratios of neighbouring
Fibonacci numbers $F_{\alpha+2}/F_{\alpha+1}$ \cite{Popk15b}, beginning with 
$z_1=2=F_3/F_2$ corresponding
to diffusion and Gaussian scaling function $f$, followed by
$z_\alpha=3/2, 5/3, 8/5, \dots$. Also the limit value of this sequence,
which is the golden mean $\phi = (1+\sqrt{5})/2$, can arise.

Which dynamical universality classes appear depends on which
diagonal elements of the mode coupling matrix vanish. A full
classification for $n=2$ is given in \cite{Spoh15,Popk15a}
and for general $n$ in \cite{Popk16}. For $n=2$ one can
have diffusion with $z=2$, and also exponents $z=3/2,5/3,\phi$. 
The dynamical exponent $z=3/2$ can describe
the KPZ universality class \cite{Halp15} 
(in which case the scaling function $f$ is the
celebrated Pr\"ahofer-Spohn function \cite{Prae04}), or a modified
KPZ universality class with unknown scaling function 
\cite{Spoh15}, or a L\'evy 
universality class \cite{Spoh15,Bern16,Popk16,Popk14a}. 
The $z=5/3$ L\'evy class
characterizes the heat mode in 
anharmonic chains \cite{Mend13,Xion16} and
one-dimensional fluids obeying 
Hamiltonian dynamics \cite{vanB12}.\footnote{The dynamical exponent $z=5/3$ 
has also been reported for heat transport in
hard-point particle gases \cite{Cipr05}, but universality for this system
has been challenged recently \cite{Xion12,Hurt16}.}
Experimental evidence for anomalous heat conduction has been found
in single multiwalled carbon and boron-nitride nanotubes at room temperature
\cite{Chan08}. 

The upshot of the mode coupling treatment of
NLFH is that {\it the dynamical
universality classes can be directly inferred from the
structure of the mode coupling matrices, which in turn
is fully determined by the stationary current-density relation 
$\vec{j}^\ast(\vec{\rho})$ for the conserved densities $\vec{\rho}$ 
of the system.}

The theory of non-linear fluctuating hydrodynamics combined with mode-coupling theory is rather robust. It relies fundamentally on the presence 
long-lived long wave-length modes which arise from the conservation
laws. 
Excluded are (i) systems that exhibit long-range
order in the stationary state, in which case
complex characteristic velocities indicative of phase separation 
\cite{Rama02,Chak16} may arise.
(ii) In systems with long-range
interactions other discrete dynamical exponents may appear, e.g., the
ballistic universality class with $z=1$ in nearest-neighbour hopping
with long-range dependence of the hopping rate \cite{Spoh99,Popk10,Popk11},
or in models with long-range jumps such as the Oslo rice pile model
with $z=10/7$ \cite{Gras16} or the raise-and-peel model \cite{Alca07}, 
also with $z=1$. (iii) Also integrable models with
non-local conservation laws might conceivably exhibit dynamical exponents
that are not Kepler ratios. However, so far there is no evidence
for such an anomaly \cite{Kund16}.

The family of height models considered here falls into neither of these
three long-range categories (i) -- (iii) and therefore one expects all 
dynamical exponents to be the Kepler ratios derived in \cite{Popk15b}. They appear
in combinations that can be derived from the mode coupling matrices
for a general number of conservation
laws following \cite{Popk16} and specifically for $n=2$ from the earlier work
\cite{Spoh15,Popk15a}. In the following we compute the mode coupling matrices
for the directed polymer model first for $n=2$ (corresponding to $d=3$) 
and then for general $n$ in 
order to work out the dynamical universality classes of the $n$ 
generalized height functions.

\subsubsection{Fluctuations in $d=3$}

In the following we apply the approach based on NLFH that we have
outlined above to the directed polymer model in three dimensions,
with the aim of identifying its universal classes 
through analysis of the mode-coupling matrix. 

When $f_1=0$ the matrices appearing in the quadratic term in the r.h.s.
of \eref{gen_eq} are the mode coupling matrices \eref{mcmatrix} introduced
above. One sees that the height variable $h_2$ has neither a 
quadratic self-coupling nor a non-linear coupling to $h_1$. 
On the other hand, $h_1$ has a non-vanishing quadratic non-linearity, 
but no quadratic coupling to the diffusive mode.
Hence according to \cite{Spoh15,Popk15a}
the evolution of $h_2$ is diffusive and mode 1 is KPZ. 

In the matrices $N$ and $P$ one recognizes the mode coupling matrices
$G^\alpha$ \eref{mcmatrix} arising from NLFH. 
Thus the universality classes can be identified.
Since both mode coupling matrices have generically non-vanishing self-coupling 
coefficients $N_{11}$ and $P_{22}$ we arrive at the conclusion that 
generically the
two-component height model
has two KPZ modes drifting away from each other with speeds \eref{lambda12}. 
Similar models were studied by Kim and den Nijs \cite{KimNijs} and Ferrari, Sasamoto and Spohn \cite{Ferr13}. 

Notice, however, that $s(f,f) = s(f,0) = f$
Therefore when 
 $f_1=f_2=:f \neq 0$ one has
$s(f,f) = f$ and therefore $N_{11}=0$, $P_{22}\neq 0$. In this case mode 1
is diffusive while mode 2, which has no coupling to the diffusive mode,
is KPZ. On the other hand, when $f_1=f\neq 0$ and $f_2=0$ one gets
$N_{11}\neq 0$, $P_{22}= 0$, which is the same scenario with the role
of two modes interchanged. Therefore also mixed dynamics may occur.
In the trivial case where $f_1=f_2=0$ both modes are diffusive.

\section{The $n$-component particle exchange process}
\label{PEP}

As discussed above the mapping between the height model and
exclusion can be applied to any dimension $d\geq 2$. Here we
define the corresponding multi-species exclusion process and discuss it in
detail in the hopping rates for which the stationary distribution factorizes.
We shall call this process the $n$-component particle exchange process (PEP).
For more general exclusion processes with nearest-neighbour
particle exchange and non-factorized stationary distributions we 
refer to \cite{Isae01,Anev03} and, for the present context, to \cite{Ferr13}.
We derive the exact mode coupling matrices in explicit form
and thus identify the possible universality classes for arbitrary 
dimension $d$.

\subsection{Definition and stationary properties}

In the $n$-component PEP an
exclusion particle of type $\alpha \in \{0,1,\dots,M\}$ on site $k$ exchanges with type $\beta$ on site $k+1$ with rate 
$g_{\alpha,\beta}$, symbolically
$$ A_\alpha A_\beta \to A_\beta A_\alpha \quad \mbox{ with rate } g_{\alpha,\beta}.$$
Type 0 is called vacancy and we speak of $M$ distinct conserved species of particles. The total
number of particles of each species in the system is denoted $N_{\alpha}$. We consider $L$ sites
with periodic boundary conditions. It is convenient to decompose the rates into a symmetric part $w_{\alpha,\beta}=w_{\beta,\alpha}>0$
for $\alpha\neq\beta$ and an 
antisymmetric part $f_{\alpha,\beta}=-f_{\beta,\alpha}$ in the form
\bel{rates}
g_{\alpha, \beta} = \frac{1}{2} (w_{\alpha, \beta} + f_{\alpha,\beta}).
\ee
Positivity of the rates implies $w_{\alpha,\beta} \geq |f_{\alpha,\beta}|$. 
For convenience we define
$w_{\alpha,\alpha}=f_{\alpha,\alpha}=0$ and denote the vacuum driving fields
for particles with vacant neighbors by
\bel{Edef}
f_\alpha := f_{\alpha,0}
\ee
which implies, by definition, $f_{0}=0$.
If for some $\alpha$ one has $w_{\alpha,0} = |f_\alpha| $, the vacuum motion of species
$\alpha$ is totally asymmetric.

From pairwise balance \cite{Schu96} we find that the canonical 
stationary distribution with $N_{\alpha}$ particles
is uniform, provided that the condition
\bel{stat}
f_{\alpha,\beta} = f_\alpha - f_\beta
\ee
is satisfied with driving fields in the physical domain  
$|f_\alpha| \leq w_{\alpha,0}$.
It follows that the grandcanonical stationary ensemble with
fluctuating particle numbers is a product measure
defined by fugacities $\mu_\alpha$, or equivalently, particle densities
\be
\rho_\alpha := \exval{N_{\alpha}}/L 
= \frac{\rme^{\mu_\alpha} }{ \sum_{\alpha'=0}^M \rme^{\mu_\alpha'}}
\ee
The product structure leads to the covariances (generalized compressibilities)
\bel{sus}
\kappa_{\alpha \beta} := \frac{\partial \rho_\alpha}{\partial \mu_\beta} =
1/L \exval{(N_{\alpha}-\exval{N_{\alpha}})(N_{\beta}-\exval{N_{\beta}})} = \rho_\alpha 
(\delta_{\alpha,\beta}-\rho_\beta).
\ee
We denote the compressibility
matrix with matrix elements $\kappa_{\alpha\beta}$ by $K$. 
By construction $K=K^T$ is symmetric.

Consider the local density $\rho_k^\alpha :=
\exval{n_k^\alpha}$, i.e., the expectation of the local particle 
number $n_k^\alpha \in \{0,1\}$. Particle number conservation implies
the discrete continuity equation
\be
\frac{d}{dt} \rho_k^\alpha = j_{k-1}^\alpha - j_k^\alpha
\ee
where, by definition of the process, the expected local current of 
species $\alpha$ is given by
\be
 j_k^\alpha = \sum_{\beta=0}^M g_{\alpha,\beta} \exval{n_k^\alpha n_{k+1}^\beta} - g_{\beta,\alpha} \exval{n_k^\beta n_{k+1}^\alpha}.
\ee
In the grandcanonical stationary distribution one has
\bel{statcur}
j_\alpha = \rho_\alpha 
\left(f_\alpha- \sum_{\beta=1}^M f_\beta \rho_\beta\right).
\ee
This follows from the  factorization property of the
grandcanonical stationary distribution.

\subsection{Collective velocities}

As dicussed above one expects 
in the hydrodynamic limit on Euler scale the system of conservation laws
\eref{hyper} where $J$ is the flux Jacobian with matrix elements
\bel{velmat}
J_{\alpha\beta} = \frac{\partial j_\alpha}{\partial \rho_\beta} = 
\left[ f_\alpha  - \sum_{\gamma=1}^M f_{\gamma} \rho_\gamma\right]  \delta_{\alpha,\beta}-  f_\beta \rho_\alpha
\ee
In order to derive the normal modes
for non-zero densities and non-zero driving fields we introduce the diagonal matrices $\hat{\rho} := \diag(\rho_\alpha)$ and 
$\hat{f}:=\diag(f_\alpha)$ with the
densities and driving fields resp. on the diagonal. Then we can write
\be
J = D^{-1} B D
\ee
where $D = \sqrt{\hat{f}/\hat{\rho}}$ and $B=B^T$. The non-diagonal matrix elements of $B$ are
$B_{\alpha\beta} = -\sqrt{f_\alpha \rho_\alpha f_\beta \rho_\beta}$. This implies that $J$ can be diagonalized 
with the help of $D$ and an orthgonal matrix $\mathcal{O}$. With $\hat{J} := \diag(v_i)$ one can write
\be
R J R^{-1} = \hat{J}
\ee
where
$R = Q^{-1} \mathcal{O} D$ and $R^{-1} = D^{-1}\mathcal{O}^T Q$ with an invertible diagonal matrix $Q = \diag(q_\alpha)$.
Notice also that $R^T = D \mathcal{O}^T Q^{-1}$ and $(R^{-1})^T = Q \mathcal{O} D^{-1}$.
Choosing $Q$ such that
\bel{ortho}
R K R^T = \mathds{1}
\ee
one obtains an orthonormal basis of the modes. To compute the matrix $Q$ we observe that
\be
J = K D^2 - \sum_\alpha f_\alpha\rho_\alpha \mathds{1}.
\ee
Thus
$RKR^T =  Q^{-1} \mathcal{O} D J D^{-1} \mathcal{O}^T Q^{-1} +
\sum_\alpha f_\alpha\rho_\alpha Q^{-1} \mathcal{O}  \mathcal{O}^T Q^{-1} = 
(RAR^{-1}  + \sum_\alpha f_\alpha\rho_\alpha \mathds{1})Q^{-2}$. This yields
\bel{q}
q_\alpha^{2} = v_\alpha + \sum_\alpha f_\alpha\rho_\alpha.
\ee

We remark that decomposing $J$ into a traceless part and the trace yields
\bel{traceless}
J = \tilde{J} + \frac{1}{M} \sum_\alpha f_\alpha \left(1-(M+1)\rho_\alpha\right) \mathds{1}
\ee
which can be written in the form $J = D^{-1} \tilde{B} D +V \mathds{1}$ with traceless and symmetric $\tilde{B}$.
For the completely symmetric state with $\rho_\alpha = 1/(M+1)$ 
as for the generalized height model one has $V=0$ and the collective velocities are the
eigenvalues of $\tilde{B}$. On the other hand, for equal driving fields $f_\alpha=f$ one has $V= f(1-\sum_{\alpha=1}^M\rho_\alpha)$
which vanishes only for the completely filled lattice. This then is the multispecies simple exclusion process.

\subsection{Mode-coupling coefficients}

The Hessians 
\bel{Hessdef}
H^{\gamma}_{\alpha\beta} := \frac{\partial^2 j^\gamma}{\partial \rho_\alpha \partial \rho_\beta} = \partial_\alpha J_{\gamma\beta}
\ee
are constants
\bel{Hess}
H^{\gamma}_{\alpha\beta} = - \left( f_\alpha \delta_{\beta,\gamma} + f_\beta \delta_{\alpha,\gamma} \right).
\ee
This simple form allows us to compute explicitly the mode-coupling coefficients 
\bel{G}
G^\gamma_{\alpha\beta} := \frac{1}{2} \sum_{\lambda} R_{\gamma \lambda} \left[ (R^{-1})^T H^{\lambda} R^{-1}\right]_{\alpha\beta}.
\ee
According to the definitions given above we have
\be
D_{\alpha\beta} = \sqrt{\frac{f_\alpha}{\rho_{\alpha}}} \delta_{\alpha,\beta}
\ee
and
\bea
R_{\alpha\beta} = q_\alpha^{-1} \sqrt{\frac{f_\beta}{\rho_{\beta}}} \mathcal{O}_{\alpha\beta}, &&
(R^{-1})_{\alpha\beta} = q_\beta \sqrt{\frac{\rho_{\alpha}}{f_\alpha}} \mathcal{O}_{\beta\alpha} 
= q^{2}_\beta \frac{\rho_{\alpha}}{f_\alpha} (R^T)_{\alpha\beta}, \nonumber \\
(R^T)_{\alpha\beta} = q_\beta^{-1} \sqrt{\frac{f_\alpha}{\rho_{\alpha}}} \mathcal{O}_{\beta\alpha}, &&
((R^{-1})^T)_{\alpha\beta} = q_\alpha \sqrt{\frac{\rho_{\beta}}{f_\beta}} \mathcal{O}_{\alpha\beta} =
q^{2}_\alpha \frac{\rho_{\beta}}{f_\beta} R_{\alpha\beta} .
\eea
Hence, by straightforward computation
\bea
G^\gamma_{\alpha\beta} & = & \frac{1}{2} \sum_\lambda \sum_\mu \sum_\nu
R_{\gamma \lambda} ((R^{-1})^T)_{\alpha\mu} H^{\lambda}_{\mu\nu} (R^{-1})_{\nu\beta} \nonumber \\
& = & -  \frac{1}{2}  \sum_\mu \sum_\nu 
\left[ f_\mu R_{\gamma \nu} ((R^{-1})^T)_{\alpha\mu}  (R^{-1})_{\nu\beta}
+ f_\nu R_{\gamma \mu} ((R^{-1})^T)_{\alpha\mu}  (R^{-1})_{\nu\beta} \right] \nonumber \\
& = & -  \frac{1}{2}  \sum_\mu \left[ f_\mu \, ((R^{-1})^T)_{\alpha\mu} \, \delta_{\beta,\gamma} +
 f_\mu \, (R^{-1})_{\mu\beta} \, \delta_{\alpha,\gamma} \right] \\
& = & -  \frac{1}{2} \left[  q^{2}_\alpha \sum_\mu R_{\alpha\mu} \rho_\mu \, \delta_{\beta,\gamma} +
q^{2}_\beta \sum_\mu R_{\beta\mu} \rho_\mu \, \delta_{\alpha,\gamma} \right] \nonumber \\
& = & -  \frac{1}{2} \left[ q^{2}_\alpha (R\vec{\rho})_\alpha \, \delta_{\beta,\gamma} + 
q^{2}_\beta (R\vec{\rho})_\beta \, \delta_{\alpha,\gamma} \right]. \nonumber
\eea
We point out the non-trivial trilinear property 
$G^\gamma_{\alpha\beta} = G^\alpha_{\gamma\beta}$ which one expects for systems
where the driving force does not change the stationary distribution 
\cite{Ferr13,Funa15}. 

For the diagonal elements one has
\be
G^\gamma_{\alpha\alpha} = - q^{2}_\alpha \, (R\vec{\rho})_\alpha \, \delta_{\alpha,\gamma}.
\ee
Hence generically all modes are KPZ and there are only subleading corrections since
$G^\gamma_{\alpha\alpha} = 0 $ for $\alpha \neq \gamma$. If one of the coefficients
$q^{2}_\alpha (R\vec{\rho})_\alpha$ vanishes, then this mode is diffusive and all other modes
evolve independently of this mode.

\subsection{Details for two conservation laws}

We return to the case $n=2$ and at least one driving field non-zero and present
the diagonalization explicitly and in detail for arbitrary densities. 
For the two-component PEP with arbitrary densities $\rho_\alpha$
we have
\be
j_1 = f_1 \rho_1(1-\rho_1) - f_2 \rho_1\rho_2, \quad j_2 = f_2 \rho_2(1-\rho_2) - f_1 \rho_1\rho_2
\ee
and the compressibility matrix is given by
\be
K =  \left( \ba{cc} \rho_{1}(1-\rho_1) & -\rho_1\rho_{2}  \\
                                    -\rho_1\rho_{2} & \rho_{2}(1-\rho_2) \ea \right).
\ee

We find
\bea
J & = & \left( \ba{cc} f_1(1-2\rho_1) - f_2 \rho_{2} & -f_2 \rho_1  \\
                                    -f_1\rho_{2} & f_2 (1-2\rho_2) -f_1\rho_{1}\ea \right) \nonumber \\
& = & \left( \ba{cc} \frac{1}{2}(f_1 (1-\rho_1) - f_2(1- \rho_{2})) & -f_2 \rho_1  \\
                                    -f_1\rho_{2} & -\frac{1}{2}(f_1 (1-\rho_1) - f_2(1- \rho_{2}))\ea \right) \\
& & + \frac{1}{2} \sum_{\alpha=1}^2 f_\alpha (1-3\rho_\alpha) \mathds{1} \nonumber .
\eea
In order to compute the eigenvalues of $J$ we use \eref{traceless}.
For $n=2$ this yields as eigenvalues of $\tilde{B}$ the quantities $\pm \sqrt{\det{\tilde{B}}}$ and therefore
\be
v_{1,2} =  \frac{1}{2} \left[ \sum_{\alpha=1}^2 f_\alpha \left(1-3\rho_\alpha\right)
\pm \sqrt{ [f_1(1-\rho_1) - f_2(1-\rho_2)]^2 + 4 f_1\rho_1 f_2\rho_2} \right].
\ee
In the domain of interest $0 < \rho_1+\rho_2 <1$ one has $\det{\tilde{B}}>0$. Hence the 
corresponding system of conservation laws is strictly hyperbolic. For the
special case $\rho_1=\rho_3=1/3$ we recover \eref{lambda12}.

In order to compute $R$ we define the orthgonal matrix
\be
\mathcal{O} = \left( \ba{cc} \cos{\phi} & -\sin{\phi} \\
                                      \sin{\phi} & \cos{\phi} \ea \right).
\ee
Straightforward computation shows that $J$ is diagonalized with the choice
\bel{2phi}
\tan{(2\phi)} = \frac{2\sqrt{f_1 \rho_1 f_2 \rho_2}}{f_1(1-\rho_1) - f_2(1-\rho_2)}.
\ee
This yields, together with \eref{ortho},
\be
R =  \left( \ba{cc} q_1^{-1} \sqrt{\frac{f_1}{\rho_{1}}}\cos{\phi} & -q_1^{-1}\sqrt{\frac{f_2}{\rho_{2}}}\sin{\phi} \\
                                    q_2^{-1}  \sqrt{\frac{f_1}{\rho_{1}}}\sin{\phi} & q_2^{-1} \sqrt{\frac{f_2}{\rho_{2}}}\cos{\phi} \ea \right),
R^{-1} =  \left( \ba{cc} q_1 \sqrt{\frac{\rho_{1}}{f_1}}\cos{\phi} & q_2\sqrt{\frac{\rho_{1}}{f_1}}\sin{\phi} \\
                                     - q_1 \sqrt{\frac{\rho_{2}}{f_2}}\sin{\phi} & q_2 \sqrt{\frac{\rho_{2}}{f_2}}\cos{\phi} \ea \right)
\ee
where
\be
q_i^{2} = v_i + f_1\rho_1 + f_2\rho_2.
\ee

The Hessians are
\be
H^{1} = -  \left( \ba{cc} 2f_1 & f_2 \\
                            f_2 & 0\ea \right), \quad 
H^2 = - \left( \ba{cc} 0 & f_1 \\
                            f_1 & 2f_2\ea \right)
\ee
and \eref{G} yields
\be
G^{1} = - \frac{1}{2} \left( \ba{cc} 2g_1 & g_2 \\
                            g_2 & 0\ea \right), \quad 
G^2 = - \frac{1}{2} \left( \ba{cc} 0 & g_1 \\
                            g_1 & 2 g_2\ea \right)
\ee
with coupling constants
\bel{g1g2}
g_1 = q_1 \left( \sqrt{f_1\rho_1} \cos{\phi} - \sqrt{f_2\rho_2} \sin{\phi}\right), \quad
g_2 = q_2 \left( \sqrt{f_1\rho_1} \sin{\phi} - \sqrt{f_2\rho_2} \cos{\phi}\right).
\ee
As expected, generically both modes are KPZ with subleading corrections.

Care has to be taken if $f_1=0$ and $f_2=f\neq0$. Then 
\be
j_1 = -f \rho_1\rho_2,  \quad j_2 = f \rho_2(1-\rho_2)
\ee
and
\be 
J = \left( \ba{cc} -f \rho_2 & -f \rho_1 \\
                            0 & f (1-2\rho_2) \ea \right).
\ee
This yields the collective velocities
\be
v_1 = -f \rho_2, \quad v_2 = f(1-2\rho_2)
\ee
and
\be 
R = \left( \ba{cc} \sqrt{\frac{1-\rho_2}{\rho_1(1-\rho_1-\rho_2)}} & \sqrt{\frac{\rho_1}{(1-\rho_1-\rho_2)(1-\rho_2)}} \\
                            0 &  \frac{1}{\sqrt{\rho_2(1-\rho_2)}} \ea \right), \quad 
R^{-1} = \left( \ba{cc} \sqrt{ \frac{\rho_1(1-\rho_1-\rho_2)}{1-\rho_2}} & - \rho_1\sqrt{ \frac{\rho_2}{1-\rho_2}} \\
                            0 & \sqrt{\rho_2(1-\rho_2)}\ea \right).
\ee
For the Hessians one has
\be 
H^{1} = - f \left( \ba{cc} 0 & 1 \\
                            1 & 0\ea \right), \quad 
H^2 = -2f \left( \ba{cc} 0 & 0 \\
                            0 & 1\ea \right).
\ee
Then \eref{G} yields
\be 
G^{1} = - \frac{f}{2} \sqrt{\rho_2(1-\rho_2)} \left( \ba{cc} 0 & 1 \\
                            1 & 0\ea \right), \quad 
G^2 = - f \sqrt{\rho_2(1-\rho_2)} \left( \ba{cc} 0 & 0 \\
                            0 & 1\ea \right).
\ee
(We remind the reader that the labels at $G$ and $H$ are upper indices, not powers). 
Hence mode 1 is diffusive and mode 2 is KPZ. 

The case $f_2=0$ and $f_1=f\neq0$ follows by symmetry. We also consider $f_1=f_2=f$. 
In this case \eref{2phi} yields
$\tan{\phi} = \sqrt{\rho_1/\rho_2}$ in which case \eref{g1g2} gives $g_1=0$, or $\tan{\phi} = -\sqrt{\rho_2/\rho_1}$ 
in which case $g_2=0$. Hence one of the modes is diffusive, as first argued in \cite{Rako04}.


\section{Conclusions}
\label{Conc}

We have treated coupled non--linear stochastic PDE equations of KPZ 
type in two different contexts: A model for directed polymers in $d=3$ where we 
derived from a dynamical mean field approach a system of two coupled partial differential equations, and from non--linear 
fluctuating hydrodynamics theory where the same equations are shown
to follow from conservation laws for the densities and the presence 
of noise.  These 
equations can then be treated in mode coupling theory. Both approaches lead to the same structure of the mode 
coupling terms. 

Next we generalized the lattice gas approach to an arbitrary
number $n$ of conserved particle species, corresponding to a model for directed
polymers in $d=n+1$ dimensions. Thus we give a direct physical link between 
fluctuations in the conformations of the polymer and the 
underlying particle exchange processes on the lattice. This allows in particular to 
understand and access different cases of the general classification given in 
\cite{Spoh15,Popk15a} for two conservation laws and for an arbitrary
number of conservation laws in \cite{Popk15b,Popk16}. It turns out that
stationary spatio-temporal fluctuations in the polymer model 
are generally either diffusive or
in the universality class of the one-dimensional KPZ equation.

There is an interesting open problem: The totally asymmetric two-component model ($w_{10}=f_1$, $w_{20}=f_2$ $w_{12}=f_1-f_2$) 
is integrable  \cite{Popk02}. Can one use the integrability to obtain directly
the exact scaling form of the dynamical structure function?
From the results of the present work one expects this to be the Praehofer-Spohn scaling function \cite{Prae04} for each mode separately.

\section*{Acknowledgements}
It is our special  honour and pleasure to thank David Huse without whom this paper
would not exist. The interesting and enlightening discussions with BK during her visit to the IAS
were instrumental in the early stages of this research. We gratefully acknowledge his sharing of ideas and intuition.

We thank DFG for financial support. BK  thankfully acknowledges support from the  NSF under the grants PHY-0969689 and PHY-1255409. She also thanks the IAS in Princeton and the Max-Planck Institute for Mathematics in Bonn for their hospitality.

\end{document}